\documentclass[aps,pra,amsmath,showpacs,twocolumn,groupedaddress,a4paper]{revtex4}

\usepackage{graphics}
\newcommand \beq{\begin{equation}}
\newcommand \beqa{\begin{eqnarray}}
\newcommand \beqann{\begin{eqnarray*}}
\newcommand \eeq{\end{equation}}
\newcommand \eeqa{\end{eqnarray}}
\newcommand \eeqann{\end{eqnarray*}}

\newcommand \la{\raisebox{-.5ex}{$\stackrel{<}{\sim}$}}
\newcommand{\ve}[1]{\mbox{\boldmath $#1$}}

\begin{document}
\title{Structure of vortices in rotating Bose-Einstein condensates}
\author{Gentaro Watanabe,$^{a,b}$ S. Andrew Gifford,$^{c}$ Gordon Baym,$^{a,c}$
and C. J. Pethick$^{a}$}
\affiliation{
$^{a}$NORDITA, Blegdamsvej 17, DK-2100 Copenhagen \O, Denmark
\\
$^{b}$The Institute of Chemical and Physical Research (RIKEN),
2-1 Hirosawa, Wako, Saitama 351-0198, Japan
\\
$^{c}$Department of Physics, University of Illinois, 1110 W. Green Street,
Urbana, IL 61801
}


\begin{abstract}

    We calculate the structure of individual vortices in rotating
Bose-Einstein condensates in a transverse harmonic trap.  Making a
Wigner-Seitz approximation for the unit cell of the vortex lattice, we derive
the Gross-Pitaevskii equation for the condensate wave function in each cell of
the lattice, including effects of varying coarse grained density.  We
calculate the Abrikosov parameter, the fractional core area, and the energy of
individual cells.

\end{abstract}


\maketitle

\section{Introduction\label{sect intro}}

    Rapid rotation of an atomic Bose-Einstein condensate has made it possible
to create vortex lattices in which the vortex core size is comparable to the
intervortex spacing \cite{first,spoon,lattice}.  This has opened up a new
regime inaccessible in superfluid He-II.  At low rotation rates, the vortex
core size is small compared with the intervortex spacing, and it is given by the
Gross-Pitaevskii healing length \cite{gross,pitaevskii,fetter_core}.  However,
as the angular velocity $\Omega$ is increased, the vortex core radii grow
and eventually become comparable to the separation of vortices 
\cite{fb,bp,coddington,schweik}.  In a harmonic trap, when $\Omega$ is close
to the transverse trapping frequency $\omega_{\perp}$, the condensate wave
function is dominated by the lowest Landau level (LLL) of the Coriolis force 
\cite{ho}.  The criterion to be in the LLL regime is that $\hbar \Omega \gg
gn$, where $gn$ is the interaction energy, $n$ is the particle density,
$g\equiv 4\pi\hbar^2 a_{\rm s}/m$ is the two-body interaction strength, $m$ is
the particle mass, and $a_{\rm s}$ is the $s$-wave scattering length. 
If the system at rest can be described by the
Gross-Pitaevskii (GP) equation for the condensate wave function, then the GP approach is valid for any rotation rate until the
number of vortices $N_{\rm v}$ becomes of order 10\,\% of the number of
particles $N$ \cite{wilkin,sinova,cooper,elast}.

    The purpose of this paper is to calculate the internal structure of the
vortices in a lattice for arbitrary rotation velocities and position in the
lattice.  Such calculations are needed to analyze current experiments in which
$gn$ is comparable to $\hbar \Omega$, and one is between the limits of slow
rotation and the LLL regime.  This work generalizes an earlier account 
\cite{bp}, which assumed that the vortex structure was the same throughout the
lattice.  It is also complementary to the work of Cozzini, Stringari, and
Tozzo \cite{cozzini}, who investigated numerically the vortex structure in the
case when the rotation rate equals the transverse trapping frequency, 
and thus the
particle density is uniform.  Our basic approach, as in Ref.  \cite{bp}, is to
treat a unit cell of the vortex lattice in the Wigner-Seitz approximation, in
which one replaces the actual unit cell, hexagonal for a triangular lattice,
by a circular one.  In Sec.~\ref{sect formalism}, we present the basic
formalism.  We find that the structure depends not only on the rotation rate
and local density, but also on derivatives of the local density (an effect
taken into account only globally in Ref.  \cite{bp}).
Numerical calculations of the local structure of vortices are given in 
Sec.~\ref{sect core}.
In Sec.~IV we calculate the Abrikosov parameter, fractional core area, and the
energy of a Wigner-Seitz cell, and discuss the spatial dependence of the
vortex structure.

\section{Basic Formalism
\label{sect formalism}}

    In this section, we derive the GP equation for a single Wigner-Seitz cell
of the vortex lattice.  We assume a separable trapping potential,
\beq
 V({\bf r})=V_\perp ({\bf r}_\perp) +V_z(z),
\eeq
where ${\bf r}_\perp$ is the coordinate perpendicular to the rotation
axis, which we take to be in the $z$ direction.  We also assume that the
confinement in the $z$ direction is sufficiently tight that the system is
essentially two-dimensional, and that the condensate wave function $\Psi$ is
separable,
\beq
\Psi=\psi({\bf r}_\perp) H(z).
\eeq
We normalize $\Psi$ to the number of particles by taking $\int d{\bf r}_\perp
|\psi|^2=N$ and $\int dz |H|^2=1$.   In the remainder of this paper we
consider only a two-dimensional system and drop the $\perp$ subscript.

    The structure of the condensate wave function in the transverse direction
is determined by the two-dimensional GP equation,
\beq
 \left(-\frac{\hbar^2}{2m}\nabla^2 +V
  +g_{\rm 2D} |\psi|^2\right)\psi=\mu \psi.
\label{gp}
\eeq
Here $\mu$ is the chemical potential, excluding the contributions from the
motion in the $z$ direction.
The effective two-dimensional interaction strength is
\beq
  g_{\rm 2D} =g \int dz |H|^4.
\eeq
For a system uniform in the $z$ direction, $g_{\rm 2D} =g/Z$, where $Z$ is
the extent of the cloud in the $z$ direction, while if $H$ is the ground-state
wave function for an oscillator potential with frequency $\omega_z $, $g_{\rm
2D}=g/\sqrt{2\pi} d_z$, where $d_z= (\hbar/m\omega_z)^{1/2}$ \cite{castin}.

    We motivate our derivation of the basic equation for the vortex structure
by focusing on the simple example of a vortex at the center of the trap.  To
solve the GP equation, subject to the boundary conditions imposed by the
presence of the other vortices, we make a Wigner-Seitz approximation to the
central cell of the vortex lattice, replacing the hexagonal unit cell by a
circle of the same area.  The radius of the Wigner-Seitz cell is
\beq
\ell=\frac{1}{\sqrt{\pi n_{\rm v}}},
\label{ell}
\eeq
where $n_{\rm v}$ is the local vortex density.  The wave function in the
central cell is then a $p$-wave state,
\beq
 \psi=\sqrt{n_0}f(r)  e^{i\phi},
\label{wavefunctioncentral}
\eeq
where $\phi$ is the azimuthal angle measured relative to the center of the
trap, and $n_0$ is the average of the particle density over the central cell.
We take the average of $|f|^2$ over the cell to be unity, $\int_j d^2r\ f^2=\pi\ell^2$.  Without loss of
generality, we may take $f$ to be real; then $f$ satisfies
\begin{eqnarray}
  -\frac{\hbar^2}{2m}\left[\frac{1}{r}\frac{\partial}{\partial
    r}
   \left(r \frac{\partial f}{\partial r}\right) -
 \frac{f}{r^2} \right]
  &+& V(r) f  \nonumber\\
  &+& g_{\rm 2D} n_0 f^3 = \mu f .
\label{gp0}
\end{eqnarray}
In order that the wave function in the central cell connect smoothly
with that in neighboring cells, we require $\partial f/\partial r=0$ at the boundary of
the Wigner-Seitz cell, $r=\ell$.  Note that the structure of the vortex
depends on the local trapping potential. In the Wigner-Seitz
approximation, the structure of the vortex on the axis of the trap
depends on the angular velocity of the trap only through the length
$\ell$ that determines the size of the unit cell.  If the symmetry of
the lattice is taken into account, there will be other terms that
depend on $\Omega$ due to the flow created by the rotating
lattice of vortices outside the
central cell, but these will be small because the true hexagonal unit
cell is well approximated by a circle.

    We now turn to the general vortex lattice.  As in Refs.\ \cite{bp,fb}, we
assume that the number of vortices is large and write the condensate wave function
as the product of the square root of a slowly varying
coarse-grained density $n$, a rapidly varying real function $f$, which
describes the local structure of vortices, and a phase factor $e^{i\Phi}$:
\begin{equation}
  \psi = \sqrt{n({\bf r})} f({\bf r})
  e^{i\Phi({\bf r})}\ .
\label{ansatz}
\end{equation}
As in the case of the central cell,
we normalize $f^2$ by setting its average over each unit cell of the
lattice to unity.

    We assume, in setting up the general expressions for the energy and
angular momentum of the vortex lattice, that the lattice is close to uniform,
with possible displacements depending only on the distance 
from the central axis,
and that the system is in equilibrium.  We then specialize, in calculating the
vortex structure, to a uniform triangular lattice.  In separate publications
we will investigate the equilibrium distortions of the lattice and calculate
the elastic constants \cite{elast-new}.

The local fluid velocity is given by
$\hbar\nabla\Phi/m$, which we write as a sum of the background flow,
$\mbox{\boldmath $v$}_{\rm b}$, plus a locally circulating flow:
\begin{equation}
  \frac{\hbar}{m}\nabla\Phi
   = \mbox{\boldmath $v$}_{\rm b} + \frac{\hbar}{m}\nabla\phi_j\ ,
\label{velocity0}
\end{equation}
where $\phi_j$ is the azimuthal angle measured with respect to the center
of the vortex in cell $j$.  This latter term accounts for the vorticity in
cell $j$.  We can write the background flow in general as $\mbox{\boldmath
$v$}_{\rm b}= (\hbar/m)\left[\nabla \Phi(R_j)+ \nabla \phi_{sj}(r)\right]$, where ${\bf
R}_j$ is the center of cell $j$ and $\phi_{sj}$ is regular in the cell 
\cite{elasto}.  Here we neglect effects due to $\phi_{sj}$.

\begin{figure}
\begin{center}\vspace{0.0cm}
\rotatebox{0}{\resizebox{6.0cm}{!}
{\includegraphics{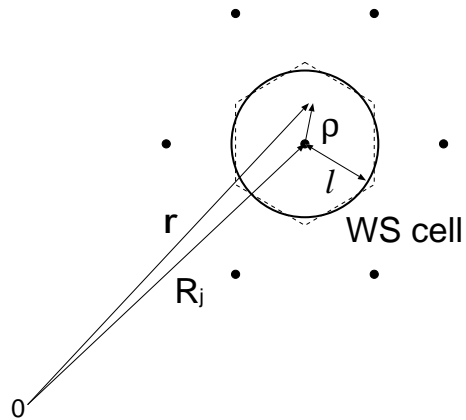}}}
\caption{\label{fig def}
A Wigner-Seitz cell of the vortex lattice.  The position of the center of
the cell $j$ relative to the center of the trap is ${\bf R}_j$, and
$\mbox{\boldmath$\rho$}$
is the transverse coordinate measured from ${\bf R}_j$.}
\end{center}
\end{figure}

    We estimate $\mbox{\boldmath $v$}_{\rm b}$ by calculating the circulation
around a circle of radius $R_j$ in terms of the number of vortices $N_{{\rm
v}j}$ within the circle:  $\mbox{\boldmath $v$}_{\rm b} = \Omega_{\rm
v}\mbox{\boldmath $\hat z$} \times {\bf R}_j$, where $\Omega_{\rm v} = \hbar
N_{{\rm v}j}/mR_j^2$.
 Thus
\begin{equation}
  \frac{\hbar}{m}\nabla\Phi =
   \ve\Omega_{\rm v}\times{\bf R}_j
  + \frac{\hbar}{m}\nabla\phi_j\ .
\label{velocity}
\end{equation}
If the lattice is uniform with vortex density $n_{\rm v}$, then
$\Omega_{\rm v} = \hbar \pi n_{\rm v}/m$.  In addition $\nabla\phi_j =
\mbox{\boldmath $\hat{\phi}$}_j /\rho$, where $\mbox{\boldmath
$\hat{\phi}$}_j$ is the unit vector in the azimuthal direction about ${\bf
R}_j$, and $\mbox{\boldmath $\rho$}= {\bf r}-{\bf R}_j$ is the 
coordinate in the $x$-$y$ plane measured from the center of the cell 
(see Fig.\ \ref{fig def}).

    We now derive the Gross-Pitaevskii equation in a Wigner-Seitz cell for a
rotating condensate in a harmonic trap.  The total energy in the laboratory
frame is [Eq.\ (8) of Ref.\ \cite{bp}]:
\begin{eqnarray}
  E &=& \int d^2r \left\{\frac{\hbar^2}{2m}|\nabla\psi|^2
   + \frac{m\omega^2 r^2}{2}\ n f^2 + \frac{g_{\rm 2D}}{2}\
  n^2 f^4\right\}
 \nonumber\\
  &\simeq&
\frac{1}{2} I\omega^2
  \nonumber\\
  && + \sum_j\int_j d^2r\ n \Biggl\{ \frac{\hbar^2}{2m}(\nabla f)^2
  +\biggl(\frac{\hbar^2}{2m}(\nabla\phi_j)^2
  \nonumber\\
  &&
  +\frac{1}{2}m\Omega_{\rm v}^2 {\bf R}_j^2
  +\hbar\mbox{\boldmath $\Omega$}_{\rm v} \cdot ({\bf R}_j\times\nabla\phi_j)
  \biggr) f^2
  +\frac{g_{\rm 2D}}{2}nf^4 \Biggr\} \nonumber\\
  &&+ \int d^2r\ \frac{\hbar^2}{2m}
  \left\{(\nabla\sqrt{n})^2 f^2+\frac{1}{2}\nabla f^2\cdot\nabla n\right\}\ ,
  \label{starter}
\end{eqnarray}
where $I = \int d^2r\ mn({\bf r})f^2 r^2$ is the
moment of inertia.  
In general, $n({\bf r})$ varies in space, and we
continue to allow for the possibility that $\Omega_{\rm v}$ depends on
position.

    The angular momentum $L$ [Eq.\ (11) of Ref.\ \cite{bp}] is similarly
\begin{eqnarray}
  L &=& \int d^2r\ nf^2 [{\bf r}\times\hbar\nabla\Phi]_z\nonumber\\
  &\simeq& \sum_j \int_j d^2r\ n f^2[m\Omega_{\rm v}(R_j^2
  + \mbox{\boldmath $\rho$}\cdot {\bf R}_j) + \hbar  \mbox{\boldmath
   $\rho$}\cdot {\bf R}_j /\rho^2]
\nonumber \\ &&+\hbar N.
\label{lws1}
\end{eqnarray}

    In the remainder of this paper we neglect the weak spatial dependence of
$\Omega_{\rm v}$ arising from the small deviations of the vortex lattice from
regular triangularity \cite{sheehy,wbp,ckr,aftalion,elast-new}, and assume
that the lattice is uniform. We also assume that $f$ is axially symmetric
within each cell.
Our objective is to calculate the energy in a frame rotating with angular velocity $\Omega$ as a functional of $f$, and then to determine $f$ by minimizing the energy. 
In Eq.\ (\ref{starter}), the energy per particle due to the trapping potential ($I\omega^2/2$) and the kinetic energy of bulk rotation (the $\Omega_{\rm  v}^2$ term) are 
of order $m \omega^2 R^2\sim \hbar (\omega^2/\Omega_{\rm v}) N_{\rm v}$ and  $m \Omega_{\rm v}^2 R^2\sim \hbar \Omega_{\rm v} N_{\rm v}$, respectively, 
where $R$ is the radius of the cloud.    The  leading contributions to these terms are independent of $f$, and therefore to determine $f$ it is necessary 
to include the leading $f$-dependent contributions, which are smaller by a factor $1/N_{\rm v}$.    The energy per particle from interparticle interactions is of order $Ng_{\rm 2D}/R^2$, and 
since this depends on $f$ it is unnecessary to include corrections of relative order $1/N_{\rm v}$.
To calculate the energy in a uniform lattice, we
expand the coarse-grained density about the center of the Wigner-Seitz cell,
$n({\bf r}) \simeq n({\bf R}_j) + \mbox{\boldmath $\rho$}\cdot\nabla
n({\bf R}_j) +\frac12(\mbox{\boldmath $\rho$}\cdot\nabla)^2 n({\bf R}_j)$.
Thus
\begin{eqnarray}
  E
  &\simeq& \frac{1}{2} I
  (\omega^2+\Omega_{\rm v}^2)
 \nonumber\\
  && + \sum_j n(R_j)\int_j d^2r\  \Biggl\{ \frac{\hbar^2}{2m}
  \left[\left(\frac{\partial f}{\partial\rho}\right)^2 + \frac{f^2}{\rho^2}
  \right]
\nonumber\\
&&
  - \frac{1}{2}m\Omega_{\rm v}^2\rho^2 f^2
  + \frac{g_{\rm 2D}}{2} n(R_j) f^4 \Biggr\}
\nonumber\\
&&
  - \frac{1}{2} \sum_j R_j \frac{\partial}{\partial R_j} n(R_j)
  \int_j d^2r\ (m\Omega_{\rm v}^2\rho^2 -\hbar\Omega_{\rm v})f^2\ .
\nonumber\\
\label{ews}
\end{eqnarray}
Here we have retained in the kinetic and trap energies 
terms of order $\hbar\Omega$ and $\hbar\omega^2/\Omega_{\rm v}$ per particle,
and thus the last two terms in Eq.\ (\ref{starter}), 
which are smaller by a factor $1/N_{\rm v}$,
are neglected \cite{neglect}.
To obtain Eq.\ (\ref{ews}), we used the fact that
\begin{eqnarray}
  &&\sum_j\int_j d^2r\ nf^2\mbox{\boldmath $\rho$}\cdot {\bf r}
\nonumber\\
  && \indent    \simeq \sum_j\left(n(R_j) + \frac12
   R_j\frac{\partial}{\partial R_j} n(R_j)\right) \int_j d^2r\ f^2\rho^2,
\nonumber\\
\end{eqnarray}
where the factor 1/2 comes from averaging in two dimensions \cite{oldcalc}.

    The angular momentum $L$ [Eq.\ (\ref{lws1})] is 
\begin{eqnarray}
  L &=& \int d^2r\ nf^2 [{\bf r}\times\hbar\nabla\Phi]_z\nonumber\\
  &\simeq& I\Omega_{\rm v} + \hbar N
  - \sum_j n(R_j) \int_j d^2r\ m\Omega_{\rm v} f^2\rho^2
\nonumber\\
  && - \frac{1}{2}\sum_j R_j\frac{\partial}{\partial R_j} n(R_j)
  \int_j d^2r\ \left(m\Omega_{\rm v}\rho^2-\hbar\right)f^2.\qquad
\label{lws}
\end{eqnarray}
We write the moment of inertia as
\begin{eqnarray}
  I &=& \bar{I} + \int d^2r\ mn(f^2 -1)r^2\nonumber\\
 &\simeq&\bar{I}
  + m \sum_j
  \Biggl(n(R_j)+ R_j\frac{\partial}{\partial R_j} n(R_j)\nonumber\\
 &&+\frac{1}{4} R_j^2 \frac{\partial^2}{\partial R_j^2} n(R_j)\Biggr)
  \int_j d^2r\ (f^2-1)\rho^2\nonumber\\
 &=& \bar{I} - \frac{\hbar N}{2\Omega_{\rm v}}
  + m \sum_j n(R_j)\int_j d^2r\ f^2\rho^2
  \nonumber\\
  &&+ m \sum_j \left(R_j\frac{\partial}{\partial R_j} n(R_j)
+\frac{1}{4} R_j^2 \frac{\partial^2}{\partial R_j^2} n(R_j)\right)\nonumber\\
&&\qquad\qquad\times\int_j d^2r\ (f^2-1)\rho^2\ ,
\label{I}
\end{eqnarray}
where $\bar{I}\equiv \int d^2r\ mn({\bf r})r^2$
is the moment of inertia calculated for the coarse-grained density.
Note that here the second derivative term
gives contributions of order $\hbar\Omega$ to the energy per particle.
In the limit of a small core, i.e., for $\hbar\Omega\ll g_{\rm 2D}n$,
$I$ reduces to $\bar{I}$.
The angular momentum written in terms of $\bar{I}$ is
\begin{eqnarray}
    L &=& \bar I \Omega_{\rm v} + \frac12 \hbar N
    + \sum_j \frac{m}{2} n(R_j) \alpha \Omega_{\rm v} \int_j d^2r\ \rho^2  f^2 \nonumber\\
  &&+\sum_j\frac{m}{4} n(R_j)\beta\Omega_{\rm v}
  \int_j d^2r\ (f^2-1)\rho^2,
\label{L}
\end{eqnarray}
where
\begin{equation}
  \alpha\equiv\frac{R_j}{n(R_j)} \frac{\partial}{\partial R_j} n(R_j)
  =\frac{\partial \ln{n(R_j)}}{\partial \ln{R_j}}\ ,
\label{alpha}
\end{equation}
and
\begin{equation}
  \beta\equiv\frac{R_j^2}{n(R_j)}\frac{\partial^2}{\partial R_j^2}n(R_j).
\label{beta}
\end{equation}

    From Eqs.\ (\ref{ews}) and (\ref{L}), the total energy $E'$ in the frame
rotating with angular velocity $\Omega$ is
\begin{align}
  E' =& E-\Omega L\nonumber\\
  =& \frac{1}{2}\left(\bar{I}-\frac{\hbar N}{2\Omega_{\rm v}}\right)
  \left[(\omega^2-\Omega^2)+(\Omega-\Omega_{\rm v})^2\right]\nonumber\\
  &-\hbar N\Omega
  -\frac{\pi\hbar^2}{4m}\left(\frac{\omega^2}
  {\Omega_{\rm v}^2}-1\right) \sum_j R_j\frac{\partial}{\partial R_j} n(R_j)
\nonumber\\
  &-\frac{\pi\hbar^2}{16m\Omega_{\rm v}^2}
\left[(\omega^2-\Omega^2)+(\Omega-\Omega_{\rm v})^2\right]
\sum_j R_j^2 \frac{\partial^2}{\partial R_j^2} n(R_j)
\nonumber\\
  &+ \sum_j E_j\ ,
\label{erot}
\end{align}
where we have collected all terms dependent on $f$ in
\begin{eqnarray}
  E_j &=& n(R_j) \int_j d^2r\ \left\{\frac{\hbar^2}{2m}
  \left[\left(\frac{\partial f}{\partial\rho}\right)^2 + \frac{f^2}{\rho^2}
  \right]
  \right.\nonumber\\
  &&\qquad\qquad\quad
  +\frac{1}{2}\ m\tilde{\omega}^2\rho^2 f^2  + \frac{g_{\rm 2D}}{2}\
   n(R_j) f^4 \Bigg\}.\qquad
\label{ej}
\end{eqnarray}
The effective trap frequency $\tilde{\omega}$ is given by 
\begin{equation}
  \tilde{\omega}^2
  \equiv \omega^2 + \alpha (\omega^2-\Omega\Omega_{\rm v})
+\frac{\beta}{4} (\omega^2+\Omega_{\rm v}^2 -2\Omega\Omega_{\rm v}).
\end{equation}
[The coefficient of the term $\rho^2 f^2$ differs from that in Eq.\
(15) of Ref.\ \cite{bp} because of the assumption made there that
$f$ is independent of position.]  The explicit form of $E$ is found by setting
$\Omega=0$ in Eq.~(\ref{erot}).
The term proportional to $\bar{I}(\omega^2-\Omega^2)$ is the energy
due to the trapping and centrifugal potentials, while the term
proportional to $\bar{I}(\Omega-\Omega_{\rm v})^2$ is the extra
kinetic energy due to the fact that when $\Omega\neq \Omega_{\rm v}$
there is bulk motion in the rotating frame.
The equilibrium value of $\Omega_{\rm v}$ is determined by
minimizing $E'$ with respect to $\Omega_{\rm v}$ keeping
$n$ and $f$ fixed, $\partial E'/\partial\Omega_{\rm v}=0$.
This condition has the form
\begin{equation}
  \frac{\partial E'}{\partial\Omega_{\rm v}}
  = \bar{I}(\Omega_{\rm v}-\Omega)
  +\frac{\partial E'_{\rm res}}{\partial\Omega_{\rm v}}=0 ,
\end{equation}
where the last term is given by 
$E'_{\rm res}=E'-(\bar{I}/2)\left[(\omega^2-\Omega^2)+(\Omega-\Omega_{\rm v})^2\right]$ and is of order $N\hbar\Omega$.
From this it follows that $(\Omega-\Omega_{\rm v})/\Omega\propto 1/N_{\rm v}$,
in agreement with earlier works \cite{sheehy,wbp,ckr}.
Since we assume that $N_{\rm v}\gg 1$, 
we shall henceforth put $\Omega_{\rm v}=\Omega$. Thus
\begin{equation}
  \tilde{\omega}^2
  = \omega^2 + \left(\alpha+\frac{\beta}{4}\right) (\omega^2-\Omega^2).
\label{omegaeff}
\end{equation}

    In the fast rotation regime, in which $\Omega\simeq\omega$, 
the correction to the trap frequency due to
the $\alpha$ and $\beta$ terms is small and
$\tilde{\omega}\simeq\omega$ for all cells.  Thus the vortex
structure does not change significantly from cell to cell.  However, at lower
rotation velocities, the position-dependent $\alpha$ and $\beta$ 
corrections and the
variation of the vortex structure from cell to cell are non-negligible.  The
frequency $\tilde{\omega}$ equals $\omega$ in the central cell
and decreases with increasing $R_j$.

As explained in earlier works, for $Na_{\rm s}/Z\gg 1-\Omega/\omega$ 
\cite{sinova_prof,bp,wbp,ckr}
(which is equivalent to the condition $N_{\rm v}\gg1$)
the global density profile is to a good approximation
a Thomas-Fermi parabola, $n(r)=n(0)(1-r^2/R^2)$ \cite{tfprof}. 
Here $R$ is the radius of the cloud 
\cite{sinova_prof,bp,wbp,ckr}, and the central density is 
\cite{tkmodes,sheehy}
\begin{equation}
  n(0) =\left(\frac{Nm\omega^2}{\pi b g_{\rm 2D} }\right)^{1/2}
  \left(1-\frac{\Omega^2}{\omega^2}\right)^{1/2}\ .
 \label{TFrot}
\end{equation}
Here $b$ is the Abrikosov parameter,
\begin{equation}
  b \equiv \frac{\langle f^4 \rangle}{\langle f^2 \rangle^2}
  = \frac{\int_j d^2r f^4}{\int_j d^2r} .
\end{equation}
In deriving the Thomas-Fermi form it has been assumed that
variations of $b$ over the cloud, which amount to at the very most 16\%,
may be neglected.
The radius of the cloud is given by
\begin{equation}
  R^2=\kappa_0' b^{1/2}\ell^2
  \left(\frac{\omega^2}{\Omega^2}-1\right)^{-1/2}\ ,
\label{tfrad}
\end{equation}
with
\begin{equation}
  \kappa_0' \equiv 2\left(\frac{Ng_{\rm 2D}m}{\pi\hbar^2}\right)^{1/2}\ .
\label{kappa0}
\end{equation}

    In the experiments of Ref.\ \cite{schweik}, $\omega_z=2\pi\times 5.3$ Hz,
and $a_{\rm s}=5.6$ nm for the triplet state of $^{87}$Rb.  The number of
atoms varies from $N \sim 10^7$ for $\Omega \alt 0.95\omega$ to $N \sim
10^5$ for $\Omega \approx 0.99\omega$, and the corresponding values of
$\kappa_0\equiv\kappa_0'/\sqrt{b}$ are of order 100 and 10, 
respectively.  We calculate below for
$\kappa_0 = 10$ and 100.

    For a Thomas-Fermi profile, $\alpha =\beta=2\left[1-n(0)/n(r)\right] =
-2r^2/(R^2-r^2)$, and \cite{inversion}
\beq
 \tilde{\omega}^2 = \omega^2
 -\frac{5r^2}{2(R^2-r^2)}(\omega^2-\Omega^2).
\eeq
Varying $E_j$ with respect to $f$, subject to the normalization condition
$\int_j d^2r\ f^2=\pi\ell^2$, we obtain the GP equation in a Wigner-Seitz
cell:
\begin{equation}
  -\frac{\hbar^2}{2m}\left[\frac{1}{\rho}\frac{\partial}{\partial\rho}
  \left(\rho \frac{\partial f}{\partial \rho}\right) - \frac{f}{\rho^2} \right]
  + \frac{1}{2}\ m{\tilde \omega}^2 \rho^2 f + g_{\rm 2D} n f^3
  = \mu f\ .
\label{wsgp}
\end{equation}
This equation, with boundary conditions $f(0)=0$ and $\partial
f(\ell)/\partial\rho=0$, determines the optimized form of $f$.  It is a
generalization of Eq.~(\ref{gp0}) to vortices away from the center of the
trap.  The equation fails for vortices close to the edge of the cloud where
the change of the coarse-grained density within the cell is large and the
expansion of $n({\bf r})$ about ${\bf R}_j$ is inadequate; the
expansion is justified provided
\begin{equation}
  \frac{1}{n(r)}
  \left|\frac{\partial n(r)}{\partial r}\right| \ell
  \sim \frac{1}{\sqrt{N_{\rm v}}} \frac{r}{R}
  \frac{1}{1-r^2/R^2} \ll 1\ ,
  \label{cond1}
\end{equation}
a condition that breaks down for the outermost single layer of vortices,
where $r\to R$.

\section{Vortex Structure\label{sect core}}

    We now investigate the vortex core structure using the framework derived
in the previous section.  
It is convenient to measure lengths in units of the effective oscillator
length $\tilde{d}\equiv\sqrt{\hbar/m\tilde{\omega}}$ [cf.  Eq.\
(\ref{omegaeff})]; the GP equation (\ref{wsgp}) becomes
\begin{equation}
  -\frac{1}{\eta}\frac{\partial}{\partial\eta}
  \left(\eta\frac{\partial f}{\partial\eta}\right) + \frac{f}{\eta^2}
  + \eta^2 f + \kappa f^3 = \Lambda f\ ,
\label{nondim-wsgp1}
\end{equation}
where $\eta\equiv \rho/\tilde{d}$, $\Lambda\equiv 2m\mu
\tilde{d}^2/\hbar^2$, and $\kappa\equiv 2mg_{\rm
2D}n\tilde{d}^2/\hbar^2$.  The scaled structure of a vortex depends
only on the local parameters $\kappa$ and (via the boundary conditions)
$\ell/\tilde{d}$.  For a Thomas-Fermi density profile,
\begin{equation}
  \kappa= \kappa_0
    \left(\frac{\omega^2-\Omega^2}{\tilde\omega^2}\right)^{1/2}
   \left(1-\frac{R_j^2}{R^2}\right).
\end{equation}

\begin{figure}
\rotatebox{0}{
\resizebox{8.2cm}{!}
{\includegraphics{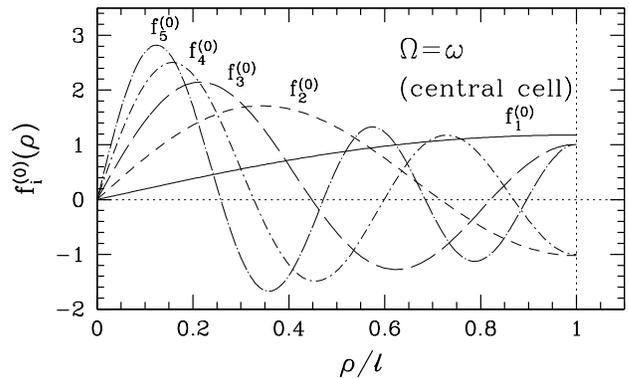}}}
\caption{\label{fig basis}
The first five basis functions $f_i^{(0)}$ of the free Hamiltonian $H_0$,
for $\Omega=\omega$ for the central cell, where
$\tilde{\omega}=\omega$.  }
\end{figure}

    We solve Eq.\ (\ref{nondim-wsgp1}) by constructing a basis set of eigenfunctions
$\{f_i^{(0)}\}$ of the free ($\kappa=0$) Hamiltonian $H_0=-\eta^{-1}
\partial_{\eta}(\eta\partial_{\eta}) + 1/\eta^2 + \eta^2$ (see, e.g., Ref.\
\cite{busch}), satisfying
\begin{equation}
  H_0 f_i^{(0)}=\Lambda_i^{(0)} f_i^{(0)}\ ,
\end{equation}
with eigenvalue $\Lambda_i^{(0)}$.  The eigenfunction $f_i^{(0)}$ of $H_0$
is
\begin{equation}
  f_i^{(0)} = A_{i}\ {}_1F_1\left(1-\Lambda_i^{(0)}/4,2;\eta^2\right)\
  \eta e^{-\eta^2/2}\ ,
\end{equation}
where $A_{i}$ is the normalization constant ensuring $\int_j
d^2\eta\ [(f_i^{(0)})^2-1]=0$, and ${}_1F_1$ is Kummer's confluent
hypergeometric function (see, e.g., Ref.\ \cite{handbook}).  The eigenvalue is
determined by the boundary condition 
\begin{equation}
  \partial_{\eta} f_i^{(0)}(\eta_{\rm cell})=0\ ,
\label{boundary}
\end{equation}
where $\eta_{\rm cell}\equiv \ell/\tilde{d} =
\sqrt{\tilde{\omega}/\Omega}$.  Figure\ \ref{fig basis} shows an example of
the first five $f_i^{(0)}$ plotted as functions of $\rho/\ell$.  The $i$th 
eigenfunction $f_i^{(0)}$ has $i-1$ nodes.  At $\Omega =\omega$ and thus
$\tilde{\omega}=\omega$, the eigenvalue $\Lambda_1^{(0)}$ of the first basis
state $f_1^{(0)}$, determined by Eq.\ (\ref{boundary}), equals 4, and
$f_1^{(0)}$ reduces to the $p$-wave function for a harmonic oscillator:
$f_1^{(0)}\sim (\rho/d)\exp{(-\rho^2/2d^2)}$.  Therefore, in this limit,
$f_1^{(0)}$ is the LLL component of the wave function in the Wigner-Seitz
cell, and the $f_i^{(0)}$, for $i\ge2$, describe the higher Landau level
components.

    We expand $f$ in Eq.\ (\ref{nondim-wsgp1}) as
\begin{equation}
  f=\sum_i c_i f_i^{(0)}\ ,
\label{expansion}
\end{equation}
and obtain the expansion coefficients $c_i$ for given $\kappa$ by
propagating the solution numerically in imaginary time, thus damping the
contribution of higher energy states.  In this analysis we use up to 15 basis
functions for $\kappa_0=10$ and up to 20 basis functions for $\kappa_0=100$.
In general the number of basis functions required increases as the core size
decreases relative to the cell size.

    Before showing results, we comment on the limits of the present analysis.
The convergence of the expansion (\ref{expansion}) becomes poor in the slow
rotation regime at larger $\kappa$, where higher energy states have
significant amplitudes.  More importantly, we have implicitly assumed [e.g.,
in Eq.\ (\ref{ansatz})] that the cloud contains a large number of vortices,
which is not the case for slow rotation.  From Eq.\ (\ref{tfrad}) one finds
\begin{equation}
 N_{\rm v}\simeq \frac{R^2}{\ell^2}
 =\kappa_0 b\left(\frac{\omega^2}{\Omega^2}-1\right)^{-1/2}\ .
\end{equation}
which is large compared with unity for $\Omega/\omega\gg
1/\sqrt{\kappa_0^2+1}$.

    In the following, we illustrate the procedure by calculating the vortex
structure close to the center of the trap, where $\alpha$ and $\beta$ can be
neglected and $\tilde{\omega}=\omega$.  Figure \ref{fig
f_omegav} shows the function $f$ obtained from the GP equation
(\ref{nondim-wsgp1}) in a Wigner-Seitz cell for
$0.5 \le\Omega/\omega\le 0.99$.  With increasing $\Omega$, the vortex
core radius increases, becoming comparable to the Wigner-Seitz cell radius at
$\Omega\agt \, 0.9\omega$.

\begin{figure}
\begin{center}\vspace{0.0cm}
\rotatebox{0}{
\resizebox{8.2cm}{!}
{\includegraphics{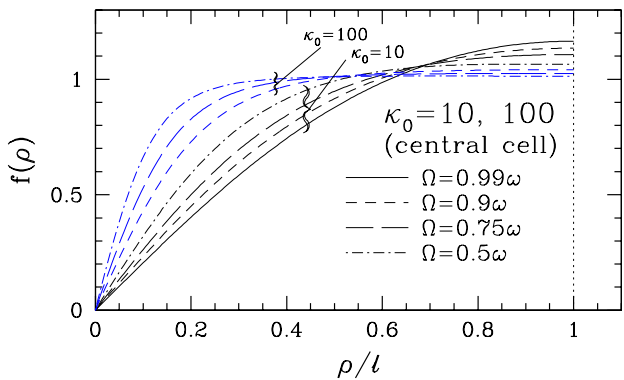}}}
\caption{\label{fig f_omegav}(Color online)\quad
  Vortex core structure at various $\Omega$ for the central cell
  obtained from the Gross-Pitaevskii equation in a Wigner-Seitz cell, with
  $\kappa_0 = 10$ (black lines) and $\kappa_0 = 100$ (blue lines).
  }
\end{center}
\end{figure}

\begin{figure}
\begin{center}\vspace{0.0cm}
\rotatebox{0}{
\resizebox{7.5cm}{!}
{\includegraphics{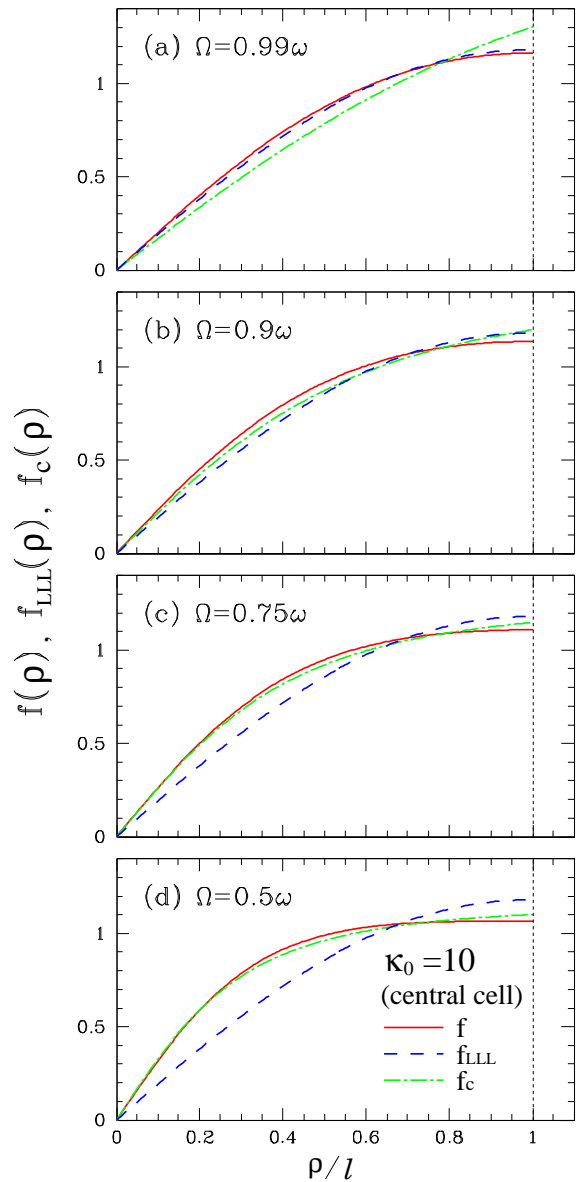}}}
\caption{\label{fig f_compare}(Color online)\quad
Comparison of the numerical solution of the local wave function $f$ with
the LLL structure $f_{\rm LLL}$ and with the single vortex form $f_{\rm c}$,
calculated in the central cell.}
\end{center}
\end{figure}

    In Fig.\ \ref{fig f_compare} we compare $f$ at $\kappa_0=10$ obtained
numerically with analytic expressions in the fast and slow rotation regimes.
In this figure
\begin{equation}
  f_{\rm LLL}(\rho) = \frac{1}{\sqrt{1-2/e}} \frac{\rho}{\ell} 
e^{-\rho^2/2 \ell^2}
\label{fLLL}
\end{equation}
is the $p$-wave function of a two-dimensional harmonic oscillator with
$d=\ell$, the lowest energy solution of Eq.\ (\ref{nondim-wsgp1})
without the interaction term in the limit $\Omega \to\omega$.

    For slow rotation, the core structure is well approximated by the single
vortex form \cite{fetter_core},
\begin{equation}
  f_{\rm c}(\rho) = A_{\rm c} \frac{\rho}{\sqrt{2\xi^2+\rho^2}}\ ,
\label{fc}
\end{equation}
where the coherence length $\xi$ is given by
\begin{equation}
  \xi^2 \equiv \frac{\hbar^2}{2mg_{\rm 2D}\ n(0)}\ ,
\end{equation}
and
\begin{equation}
  A_{\rm c} =
  \left[1+2\zeta \ln{\left(\frac{2\zeta}{1+2\zeta}\right)}\right]^{-1/2}\ ,
\end{equation}
with $\zeta\equiv\xi^2/\ell^2$, normalizes $f_{\rm c}$ 
according to $\int_j d^2r\ f_{\rm c}(\rho)^2=\pi\ell^2$.

    As Fig.\ \ref{fig f_compare}(a) shows, the numerical solution of $f$ at
$\Omega=0.99\omega$ is well described by $f_{\rm LLL}$.
However, the numerical solution starts to deviate from $f_{\rm LLL}$ with
decreasing $\Omega$ and, at $\Omega=0.9\omega$, it is
better approximated by $f_{\rm c}$ than by $f_{\rm LLL}$. For slow rotation the
single vortex form $f_{\rm c}$ is a good approximation [Figs.\ \ref{fig
f_compare}(c) and \ref{fig f_compare}(d)].

\begin{figure}
\begin{center}\vspace{0.0cm}
\rotatebox{0}{
\resizebox{8.5cm}{!}
{\includegraphics{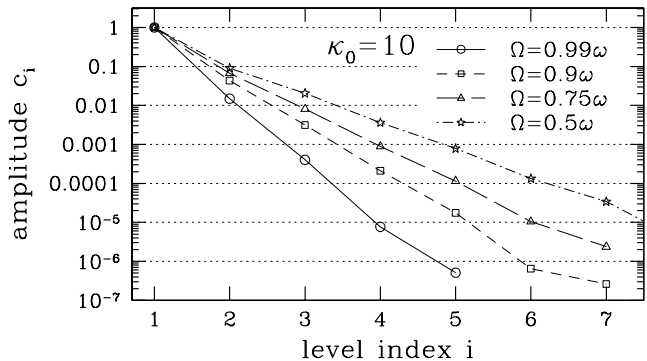}}}
\caption{\label{fig amp}
The amplitudes $c_i$ of the states $f_i^{(0)}$ in the expansion of $f$,
for the cases of $\kappa_0=10$ shown in Figs.\ \ref{fig f_omegav}
and \ref{fig f_compare}.  In
all cases here, $c_1>0.995$ and $c_i<0.1$ for $i\ge2$.
  }
\end{center}
\end{figure}

    Next we investigate the amplitude $c_i$ of the $i$th basis state
$f_i^{(0)}$ in $f$, Eq.\ (\ref{expansion}).  As Fig.\ \ref{fig amp}, a plot
of $c_i$ for each of the cases in Figs.\ \ref{fig f_omegav} and \ref{fig
f_compare} vs $i$, shows, $c_i$ decreases with $i$ almost exponentially,
$\sim\exp{[-c(i-1)]}$, where $c$ is determined by $\kappa_0$ and $\Omega$.
At $\Omega=0.9\omega$, where the system is not yet in the LLL regime,
the amplitudes for $i \ge 2$, corresponding to higher Landau levels (LL) in
the rapidly rotating limit, are less than 5\% ($c_2 \simeq 0.044$); even at
$\Omega=0.5\omega$, they are less than 10\% ($c_2 \simeq 0.090$), so
that the population in higher LL is less than 1\%.  A small admixture of order
1\% of the higher LL components is sufficient to change the internal vortex
structure from the LLL to the form for an isolated vortex.

\section{Equilibrium properties of vortices\label{sect c1}}

    We now calculate the Abrikosov parameter $b$, the fractional area of a
vortex core, and the energy of a cell of the vortex lattice, focusing on the
central cell.  In the final part of this section we discuss the spatial
dependence of the vortex structure.

\subsection{Abrikosov parameter}

\begin{figure}
\rotatebox{0}{
\resizebox{8.2cm}{!}
{\includegraphics{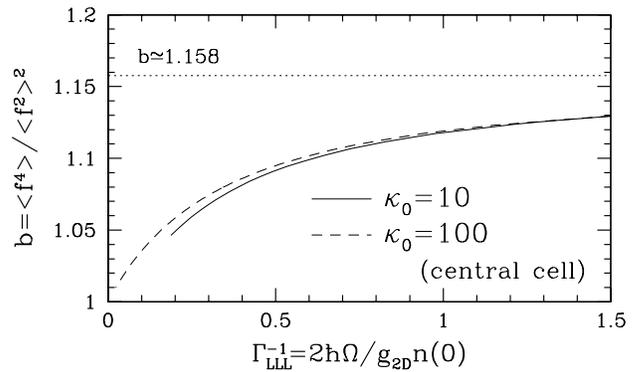}}} \caption{\label{fig abrikosov}
Abrikosov parameter $b$ vs $\Gamma_{\rm LLL}^{-1}\equiv
2\hbar\Omega/g_{\rm 2D}n(0)$, for $\kappa_0=10$ (for $\Omega \ge 0.425\omega$)
and $\kappa_0=100$ (for $\Omega \ge 0.5\omega$).
}
\end{figure}

    Figure\ \ref{fig abrikosov} shows the Abrikosov parameter as a function of
$\Gamma_{\rm LLL}^{-1}\equiv 2\hbar\Omega/g_{\rm 2D}n(0)$, while Table
\ref{values} gives specific values of $b$ for the $\Omega$ values in Figs.\
\ref{fig f_omegav} and \ref{fig f_compare}.  In the small core limit, where
$\Gamma_{\rm LLL}^{-1}\to 0$, $b=1$.  Extrapolation of the curve in Fig.\
\ref{fig abrikosov} to $\Gamma_{\rm LLL}^{-1}=0$, indicates a linear increase
in $b$ at small $\Gamma_{\rm LLL}^{-1}$, in agreement with Ref.\ \cite{fb}.
The curve increases monotonically with $\Gamma_{\rm LLL}^{-1}$ towards 
$b=\int_j
d^2r\ f_{\rm LLL}^4 /\int_j d^2r=(e^2-5)/[4(e-2)^{2}] \simeq1.158$ as $\Omega
\to \omega$ and thus $\Gamma_{\rm LLL}^{-1} \to \infty$.  This value of $b$ is
remarkably close to that for the exact LLL wave function for a triangular
lattice, $b=1.1596$ \cite{kleiner}.

\begin{table}
    \caption{Abrikosov parameter $b$ and two measures of the fractional area
of the core, $r_{\rm c}^2/\ell^2$ and $r_{\rm c}'^2/\ell^2$, at $\kappa_0=10$,
calculated for the wave functions $f$ in Figs.\ \ref{fig f_omegav} and
\ref{fig f_compare}.  }
$$
\begin{array}{ccccccc}
\hline\hline
&\Omega/\omega &\quad
\Gamma_{\rm LLL}^{-1}\quad &\quad
b\quad &\quad r_{\rm c}^2/\ell^2\quad &\quad r_{\rm c}'^2/\ell^2\quad\\
\hline
&  1.00  &  \infty &  1.158  &  0.225  &  0.368  &\\
&  0.99  &  2.807  &  1.141  &  0.217  &  0.316  &\\
&  0.90  &  0.826  &  1.115  &  0.199  &  0.231  &\\
&  0.75  &  0.454  &  1.087  &  0.177  &  0.171  &\\
&  0.50  &  0.231  &  1.056  &  0.117  &  0.103  &\\
\hline\hline
\end{array}
$$
\label{values}
\end{table}

\subsection{Fractional core area}

    Reference \cite{bp} introduced the measure of the fraction of the  area
of a Wigner-Seitz cell taken up by the vortex core:
\begin{equation}
  r_{\rm c}^2 \equiv \frac{\int_j d^2r\ [f(\ell)^2 - f(\rho)^2]\rho^2}
  {\int_j d^2r\ [f(\ell)^2 - f(\rho)^2]}\ .
  \label{rc}
\end{equation}

    The factor $\rho^2$ makes this quantity sensitive to the behavior of
$f(\rho)$ in the outer parts of the cell; in the slow rotation regime, the
slow variation of $f$ in the outer part of the cell contributes significantly
to $r_{\rm c}^2/\ell^2$.  In addition, for slow rotation (for $\Omega\la
0.4\omega$ in the present calculations) $f$ develops a local maximum for
$\rho<\ell$, which leads to a rapid decrease of $r_{\rm c}^2/\ell^2$ with
decreasing $\Omega$ at $\Gamma_{\rm LLL}^{-1} \alt 0.2$ for $\kappa_0=10$ and
at $\Gamma_{\rm LLL}^{-1} \alt 0.05$ for $\kappa_0=100$.  
The measure (\ref{rc}) of fractional area is also relatively insensitive to the
variation of the vortex wave function $f$ close to the center of the cell.

    An alternative measure of the core size which more accurately
characterizes the structure close to the vortex is
\begin{equation}
  r'_{\rm c}\equiv f(\ell)
  \left(\frac{\partial f(0)}{\partial\rho}\right)^{-1}\ .
  \label{rcdash}
\end{equation}
In addition, this quantity is not influenced by the behavior of $f$ in
the outer parts of a lattice cell.

\begin{figure}
\begin{center}\vspace{0.0cm}
\rotatebox{0}{
\resizebox{7.5cm}{!}
{
\includegraphics{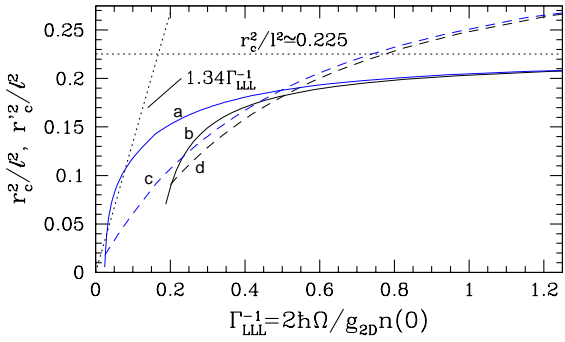}}}
    \caption{\label{fig rcsq}(Color online)\quad Fractional area of the core
for the central cell, measured by $r_{\rm c}^2/\ell^2$ (solid lines) and
$r_{\rm c}'^{2}/\ell^2$ (dashed lines) (see text), as functions of
$\Gamma_{\rm LLL}^{-1}$.  The lines a and c are for $\kappa_0=100$ and the
lines b and d for $\kappa_0=10$.  The dotted lines show the empirical
expression for the fractional area, $1.34\Gamma_{\rm LLL}^{-1}$, obtained for
low $\Omega$ \cite{coddington} and the asymptotic value of $r_{\rm
c}^2/\ell^2$ for large $\Gamma_{\rm LLL}^{-1}$ \cite{bp}.}
\end{center}
\end{figure}

    Figure\ \ref{fig rcsq} shows these two measures of the fractional area of
the core for $\kappa_0=10$ and 100 vs increasing rotation rate.  For
slow rotation, the measure $r_{\rm c}'$ more accurately describes the core
size extracted from the data \cite{schweik,coddington}.  In the rapidly
rotating regime, where the vortex core radius is comparable to the intervortex
distance, the rms radius of the density deficit $r_{\rm c}$ is readily
accessible to experiment.  For $\Gamma_{\rm LLL}^{-1} \agt 0.3$, the
calculated values of $r_{\rm c}$ are consistent with the measurements of
Refs.\ \cite{schweik,coddington} (see also Table \ref{values}).  As
$\Gamma_{\rm LLL}^{-1} \rightarrow \infty$, $r_{\rm c}^2/\ell^2$ approaches
\begin{equation}
  \frac{r_{\rm c}^2}{\ell^2}
  =\frac{11-4e}{6-2e}\simeq0.225\ ,
\end{equation}
(cf. $r_{\rm c}'^2/\ell^2=e^{-1}\simeq0.368$ as $\Gamma_{\rm LLL}^{-1}
\to \infty$).

The theoretical estimates of fractional core area are in qualitative
agreement with the published experimental data 
\cite{schweik,coddington}.  However, it is important to observe that
in the analysis of the experiments, core areas were determined from a
Gaussian fit to the profiles of the density deficit in the core.  In
addition, in our calculations we have not allowed for the
three-dimensional nature of the clouds and the effects of expansion on
vortex structure.  For these reasons, and because of uncertainties in
the number of particles in the experiments, it is premature to perform
a detailed comparison between theory and experiment.

\subsection{Energy of the central  cell}

\begin{figure}[htbp]
\begin{center}\vspace{0.0cm}
\rotatebox{0}{
\resizebox{7.8cm}{!}
{\includegraphics{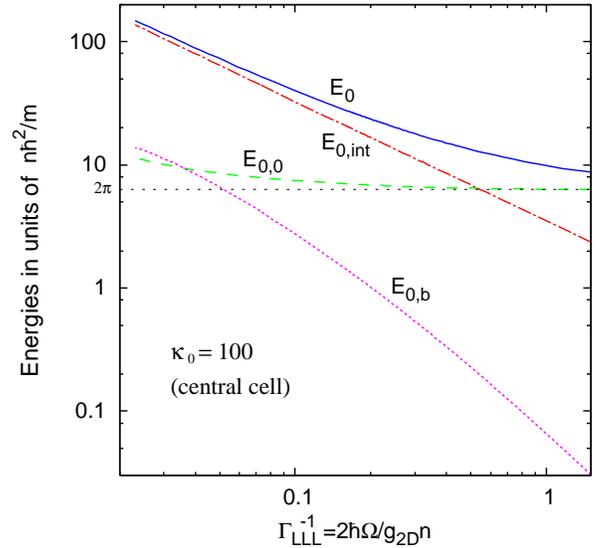}}}
\caption{\label{fig energies}(Color online)\quad
    The energy of the central Wigner-Seitz cell $E_0$ (solid line) in
units of $n\hbar^2/m$.  The kinetic plus trapping potential
energy $E_{0,0}$ (dashed line) and interaction energy $E_{0,{\rm int}}$
(dash-dotted line) are also plotted.  (In the non-interacting case,
$E_0=2\pi$.)  The ``binding'' energy $E_{0,{\rm b}}$ is the energy gain
compared to the energy for the first basis state $f_1^{(0)}$.
  }
\end{center}
\end{figure}

    In Fig.\ \ref{fig energies}, we show for the central Wigner-Seitz cell,
the energy $E_{j=0}$, the contributions from the kinetic and trap energies,
$E_{0,0}\equiv\int_{j=0} d^2r\ n\{(\hbar^2/2m)[(\partial_\rho
f)^2+f^2/\rho^2]+m\omega^2\rho^2 f^2/2\}$, and the interaction energy
$E_{0,{\rm int}}\equiv (g_{\rm 2D}/2) \int_{j=0} d^2r\ n^2f^4$, calculated with
the numerical solutions for $f$ as a function of $\Gamma_{\rm LLL}^{-1}$.  The
dependence of $E_{0,0}$ on $\Gamma_{\rm LLL}^{-1}$ is small, while $E_{0,{\rm
int}} \sim \Gamma_{\rm LLL}$.  Thus in the slow rotation limit ($\Gamma_{\rm
LLL}^{-1}\ll 1$), where the interaction energy dominates, $E_0$ also roughly
scales as $\Gamma_{\rm LLL}$.  At $\Gamma_{\rm LLL}^{-1}\simeq 0.54$,
$E_{0,0}=E_{0,{\rm int}}$ and $E_{0,0}$ becomes dominant with increasing
$\Gamma_{\rm LLL}^{-1}$.  (In the non-interacting limit with $\Gamma_{\rm
LLL}^{-1}\rightarrow \infty$, $E_0=E_{0,0}\simeq 2\pi n\hbar^2/m$.)

    To exhibit the effect of higher LL components for slower rotation, we also
show $E_{0,{\rm b}}$, the difference between $E_{0}$ and the energy calculated for the
first basis state $f_1^{(0)}$.  At $\Gamma_{\rm LLL}^{-1}=0.1$, $E_{0,{\rm
b}}$ is $\sim 7\%$ of $E_0$, falling to $\sim 0.7\%$ at $\Gamma_{\rm LLL}^{-1}=1$ (at $\Omega=0.9\omega$ and $\kappa_0=100$, $\Gamma_{\rm
LLL}^{-1}=0.0826$).  For $\kappa_0=10$, $E_0$, $E_{0,0}$, $E_{0,{\rm int}}$,
and $E_{0,{\rm b}}$ do not differ significantly from their values for 
$\kappa_0=100$ at $\Gamma_{\rm LLL}^{-1}\agt 1$.  For smaller $\Gamma_{\rm
LLL}^{-1}$, both $E_{0,{\rm b}}$ and $E_{0,{\rm b}}/E_0$ decrease with
decreasing $\kappa_0$.

\subsection{Spatial variation of vortex structure}

\begin{figure}[tbp]
\begin{center}\vspace{0.cm}
\rotatebox{0}{
\resizebox{8.2cm}{!}
{\includegraphics{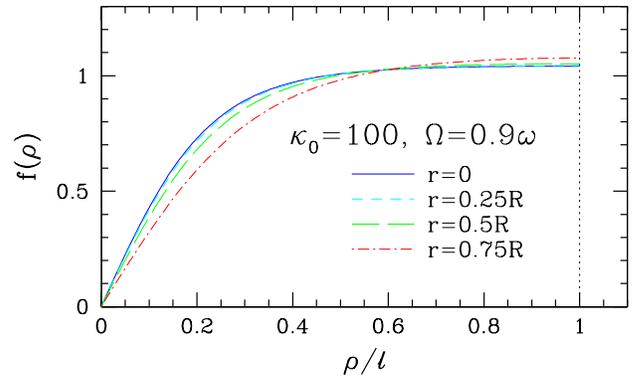}}}
\caption{\label{fig f_r}(Color online)\quad Spatial dependence of the
vortex core structure for $\kappa_0 = 100$ and $\Omega=0.9\omega$.  }
\end{center}
\end{figure}

    Although we have looked so far only at the central cell, two effects give
rise to spatial dependence of the vortex structure.  The first is that the
coarse grained density $n$ depends on position; also the effective trapping
frequency $\tilde{\omega}$ depends on spatial derivatives of $n(r)$.  Both $n$
and $\tilde{\omega}$ decrease with increasing distance from the center of the
cloud, and this increases the core radius in the outer region of the
cloud.  Figure \ref{fig f_r} shows the variation of the vortex core structure
with increasing distance $r$ from the center.  At $r=0.75R$ the core radius is
$\sim 30$\% greater than for the central cell.
\linebreak

\section{Outlook\label{summary}}

In this paper we have considered the equilibrium structure of vortices 
in two-dimensional situations.  In order to be able to make detailed 
comparison with experimental data, it is necessary to take into account the 
three-dimensional nature of the system and the expansion of the cloud
prior to the measurement of vortex structure.

    In the numerical calculation in this paper we have assumed a uniform
triangular lattice.  The next step is to extend the methods developed here to
consider distortions of the vortex lattice, to calculate the elastic
properties of the lattice, including the elastic constants $C_1$ and $C_2$ 
\cite{sinova,bc,tkmodes,sonin,cozzini}, for general rotation rates in the
presence of a spatially varying coarse grained density.

\begin{acknowledgements}
    We thank H. M. Nilsen, L. M. Jensen, and H. Smith for stimulating and
helpful discussions.  This work was supported in part by Grants-in-Aid for
Scientific Research provided by the Ministry of Education, Culture, Sports,
Science and Technology through Research Grant No. 14-7939, by the Nishina
Memorial Foundation, and by the JSPS Postdoctoral Program for Research
Abroad.  This research was also supported in part by NSF Grants No. PHY03-55014
and No. PHY05-00914.
\end{acknowledgements}


\begin{thebibliography}{99}

    \bibitem{first} M. R. Matthews, B. P. Anderson, P. C. Haljan, D. S. Hall,
C. E. Wieman, and E. A. Cornell, Phys.\ Rev.\ Lett.\ {\bf 83}, 2498 (1999).

    \bibitem{spoon} K. W.  Madison, F. Chevy, W. Wohlleben, and J. Dalibard,
Phys.\ Rev.\ Lett.\ {\bf 84}, 806 (2000).

    \bibitem{lattice} J. R. Abo-Shaeer, C. Raman, J. M. Vogels, and W.
Ketterle, Science {\bf 292}, 476 (2001).

    \bibitem{gross} E. P. Gross, Nuovo Cimento {\bf 20}, 454 (1961).

    \bibitem{pitaevskii} L. P. Pitaevskii, Zh.\ Eksp.\ Teor.\ Fiz.\ {\bf 40},
646 (1961) [Sov.\ Phys.\ JETP {\bf 13}, 451 (1961)].

    \bibitem{fetter_core} A. L. Fetter, in {\it Lectures in Theoretical
Physics}, edited by K. Mahanthappa and W. E. Brittin (New York, 1969), Vol.  XIB,
p. 351.

    \bibitem{fb} U. R. Fischer and G. Baym, Phys.\ Rev.\ Lett.\ {\bf 90},
140402 (2003).

    \bibitem{bp} G. Baym and C. J. Pethick, Phys.\ Rev.\ A {\bf 69}, 043619
(2004).

    \bibitem{coddington} I. Coddington, P. C. Haljan, P. Engels, V.
Schweikhard, S. Tung, and E. A. Cornell, Phys.\ Rev.\ A {\bf 70}, 063607
(2004).

    \bibitem{schweik} V. Schweikhard, I. Coddington, P. Engels, V. P.
Mogendorff, and E. A. Cornell, Phys.\ Rev.\ Lett.\ {\bf 92}, 040404 (2004).

    \bibitem{ho} T.-L.  Ho, Phys.\ Rev.\ Lett.\ {\bf 87}, 060403 (2001).

    \bibitem{wilkin} N. K. Wilkin, J. M. F. Gunn, and R. A. Smith, Phys.\
Rev.\ Lett.\ {\bf 80}, 2265 (1998).

    \bibitem{sinova} J. Sinova, C. B. Hanna, and A. H. MacDonald, Phys.\ Rev.\
Lett.\ {\bf 89}, 030403 (2002).

    \bibitem{cooper} N. R. Cooper, N. K. Wilkin, and J. M. F. Gunn, Phys.\
Rev.\ Lett.\ {\bf 87}, 120405 (2001).

    \bibitem{elast} G. Baym, Phys.\ Rev.\ A {\bf 69}, 043618 (2004).

    \bibitem{cozzini} M. Cozzini, S. Stringari, and C. Tozzo, Phys.  Rev.
{\bf A} 73, 023615 (2006).

    \bibitem{castin} Y. Castin and R. Dum, Eur.\ Phys.\ J.\ D {\bf 7}, 399 (1999).
   

    \bibitem{elast-new} G. Baym, S. A.  Gifford, C. J.  Pethick, and 
G. Watanabe, Phys.\ Rev. A {\bf 75}, 013602 (2007).

    \bibitem{elasto} To make contact between the present decomposition and the
long-wavelength average in the elastohydrodynamics of
Refs.~\cite{bc,elast,tkmodes},
we note that the long-wavelength average, denoted by $\langle ...  \rangle$,
of $\nabla\Phi$ is $m \ve{v}({\bf r})/\hbar$, where $\ve v({\bf r})$ is the long wavelength flow
velocity in the lab.  The velocity $\ve v_R({\bf r})$ in  a frame rotating 
at angular velocity $\Omega$
is given by $  \ve v_R({\bf r}) =  \ve v({\bf r}) - \ve \Omega \times {\bf r}$.  
This is related to the 
function $\ve \epsilon({\bf r})$  which gives the smoothed displacement of a vortex 
from its position for a uniform lattice with vortex density  
$n_{\rm v}^0=m\Omega/\pi \hbar$  by $\ve v_R +
2\ve\Omega\times \ve\epsilon({\bf r}) = (\hbar/m)\nabla \phi_s$, where $\phi_s$
describes vorticity-free superfluid flow.  Thus $\langle\nabla\Phi({\bf r})\rangle =
(m/\hbar)\ve \Omega\times \left[{\bf r}-2\ve \epsilon({\bf r})\right] +\nabla\phi_s({\bf r}),$ from
which we identify the long-wavelength limit of $\phi_{sj}({\bf r})$ with the
potential $\phi_{s}({\bf r})$.  The curl of this equation implies $\nabla\times
\langle\nabla\Phi({\bf r})\rangle = 2(m/\hbar)\ve\Omega(1-\nabla\cdot\ve\epsilon)$; since the
mean density of vortices is given by $n_{\rm v}({\bf r}) = n_{\rm v}^0 - \nabla\cdot
\left[n_{\rm v}^0\ve\epsilon({\bf r})\right]$, we
see that $ \nabla\times \langle\nabla\Phi({\bf r})\rangle = 2\pi n_{\rm v}({\bf r})$, as
expected. Note that this result may also be obtained by working in terms of a function 
giving the smoothed displacement of a vortex from the {\it equilibrium} lattice, 
which has a non-uniform density of vortices.

    \bibitem{sheehy} D. E. Sheehy and L. Radzihovsky, Phys.\ Rev.\ A {\bf 70},
051602(R) (2004); {\bf 70}, 063620 (2004).

    \bibitem{wbp} G. Watanabe, G. Baym, and C. J. Pethick, Phys.\ Rev.\ Lett.\
{\bf 93}, 190401 (2004).

    \bibitem{ckr} N. R. Cooper, S. Komineas, and N. Read, Phys.\ Rev.\ A {\bf
70}, 033604 (2004).

    \bibitem{aftalion} A. Aftalion, X. Blanc, and J. Dalibard, Phys.\ Rev.\ A
{\bf 71}, 023611 (2005).

    \bibitem{neglect} Each of these terms is of order $\hbar^2/2mR^2$, 
and thus they scale in the same way with $R$ as does the interaction energy 
$\sim g_{\rm 2D} N/R^2$.  They are thus negligible compared with 
the interaction energy provided $Na_{\rm s}/Z \gg 1$.
For $1 \agt Na_{\rm s}/Z \gg 1-\Omega/\omega$, the term
$(\hbar^2/2m) (\nabla \sqrt{n})^2 f^2$ would lead to a Gaussian 
density profile if the vortex lattice were forced to be perfect \cite{bp}.
However, 
as has been demonstrated earlier, when distortion of the vortex lattice 
is allowed for, these  contributions are largely canceled by 
contributions to the elastic energy of the vortex lattice 
and the density profile has the Thomas-Fermi form provided $N_{\rm v}\gg 1$ 
\cite{wbp}.
Within the contex of the present paper, the Thomas-Fermi form
for $Na_{\rm s}/Z\gg 1$ follows from the condition 
$\delta E'/\delta n=\mu$, where $\mu$ is the chemical potential
in the rotating frame.


    \bibitem{oldcalc} By contrast, the calculation of Ref.\ \cite{bp} took
into account only the global average of this term, which vanishes when the
density vanishes at large distances; as a consequence the present equations
differ from those of \cite{bp}, e.g., the signs of the second and third terms
of Eq.\ (\ref{I}) are different from those of Eq.\ (13) of Ref.\ \cite{bp} for the
global average.

    \bibitem{sinova_prof} J. Sinova, C. B. Hanna, and A. H. MacDonald, Phys.\
Rev.\ Lett.\ {\bf 90}, 120401 (2003).

    \bibitem{tfprof} In the present formalism the Thomas-Fermi form 
for $Na_{\rm s}/Z \gg 1$ 
follows from the condition $\delta E'/\delta n(r)= {\rm const}$ 
with $E'$ given by Eq.\ (\ref{erot}).
To establish this result within the present framework for
$1 \agt Na_{\rm s}/Z \gg 1-\Omega/\omega$, it is necessary to
allow for distortion of the vortex lattice.


    \bibitem{tkmodes} G. Baym, Phys.\ Rev.\ Lett.\ {\bf 91}, 110402 (2003).

    \bibitem{inversion} Note that the
effective oscillator potential within a cell becomes inverted when $n(r)/n(0)<
5(1-\chi)/(7-5\chi)$, where $\chi=\Omega^2/ \omega^2$; however, this
inversion, which at the rotations of interest occurs only in the outer layers
of the lattice, has little practical effect on the vortex structure.

    \bibitem{busch} T. Busch, B.-G.  Englert, K. Rz\c{a}\.{z}ewski, and M.
Wilkens, Found.\ Phys.\ {\bf 28}, 549 (1998).

    \bibitem{handbook} {\it Handbook of Mathematical Functions with Formulas, Graphs, and Mathematical Tables},
edited by M. Abramowitz and I. A. Stegun 
(National Bureau of Standards, Washington, DC, 1972), Chap. 13, p. 504.

    \bibitem{kleiner} W. H. Kleiner, L. M. Roth, and S. H. Autler, Phys.\
Rev.\ {\bf 133}, A1226 (1964).

    \bibitem{bc} G. Baym and E. Chandler, J.\ Low Temp.\ Phys.\ {\bf 50}, 57
(1983).

    \bibitem{sonin} E. B. Sonin, Phys.\  Rev.\  A {\bf 72}, 021606(R) (2005).

\end{thebibliography}
\end{document}